# Telecommunications and Rural Economies:  Findings from the Appalachian Region

Sharon Strover, Michael Oden and Nobuya Inagaki
University of Texas

ABSTRACT

Uneven access and capacity underscore the primary challenges rural communities face in exploiting the new technologies. They must secure cost- and quality-competitive access to advanced telecom services and rapidly build local expertise, training and service capacities to improve local business performance and to attract new firms.

The research strategy used here provides a comprehensive map of current telecom infrastructure patterns and focuses on tracing outcomes associated with federal and state universal service programs as well as additional state and local telecommunications-related initiatives.  This work is part of a larger project that used field research and telephone interviews, archival and secondary documents, and web-based investigations in order to gather data.

Our goal is to document the status of telecommunications in the Appalachian region with a view to assessing its potential relationship to economic growth and the range of federal and state policies that influence its development.  We find that telecommunications infrastructure in the Appalachian regions is less developed than that in other parts of the country and that it compares negatively to national averages on various broadband indicators.  Broadband technologies such as cable modems, DSL, and even the presence of high-speed services are not as widely distributed in our target region as national statistics would suggest.  Statistical analyses show that these distribution patterns are in each case associated with economic activity:  more distressed counties have less developed broadband telecommunications infrastructure.

We find that federal universal service supports favor the most rural of the Appalachian states: only Mississippi, Alabama, West Virginia, South Carolina, Georgia and Kentucky have a net positive inflow of funds through the program, although the internal adjustments (from larger, urban-serving companies to smaller, rural companies) among the other states are not to be discounted.  These six states are among the most rural of all the Appalachian states, having the lowest population densities among the group we are examining (Tennessee being a close exception).

While state universal service programs have cropped up in part to ameliorate the revenues losses local exchange companies face attribute to deregulation (especially reduced access rates), those programs are not uniform.  Most offer some low-income support as well as support to telecommunications companies serving high cost territories.  Some states are not allowing that support to flow to the largest, wealthiest companies (e.g., the BOCs or other price-cap companies) and instead favor companies serving exclusively rural regions.  In such approaches they hint at the sorts of concerns for balancing costs and supports that will probably become more pervasive in the future.

Several states have proactively initiated programs to enhance telecommunications infrastructure.  By using state telecommunications networks through resource sharing, demand aggregation or anchor tenancy programs, states are able to leverage their considerable investment and offer benefits to other public sector users - and in some cases, even private sector users.  Seven states also allow municipally owned utilities to offer telecommunications services, expanding the range of choices and the potential for competition in the process.  Nearly every state had some special program, or many programs, for enhancing Internet connectivity or broadband access.  The least active states appear to be West Virginia, Mississippi, South Carolina, Kentucky and Alabama, although these too have some state programs to enhance telecommunications access or use.

One factor that appears to enhance state potentials for improved telecommunications is coordination among state agencies within the state. By coordinating network design and use, state-funded infrastructure can be used optimally. When it is absent, programs may be duplicative, underutilized, and more costly.

Most state and federal programs have focused on market-related initiatives to solve their telecommunications problems. We observe, however, that attempting to work with (or against) the market yields only limited returns in the absence of leadership. With more creative collaboration and attention to some of the nonmarket solutions to obtaining and using telecommunications - solutions such as training, education, organizational resource sharing - the larger harnessing of telecommunications capabilities to economic growth can be enhanced.

I. INTRODUCTION

Many scholars recognize that information and telecommunications industries have become the critical drivers of the U.S. economy. These industries have had a dominant influence on recent growth performance due to their direct contribution to output and employment and through their pervasive impacts on industries and households that use their products and services (U.S. Department of Commerce, 2000). Several studies emphasize the potential benefits that new information technology could bring to rural or distressed areas by reducing the importance of market proximity and transportation costs in business location (Williams, 1991; Parker et. al, 1989, 1995). However, like earlier key technologies, the integrated architecture of computing and telecommunications exhibits a clear pattern of uneven distribution. Population density, income, geographic location, and the initial presence of innovative producers are among the main factors that influence production and use of new appliances and software systems and access to high-speed broadband networks (National Telecommunications and Information Administration, 1999; U.S. Department of Agriculture, 2000). These factors, affecting both access and capacity to use advanced telecom technologies, suggest that poorer rural regions actually risk falling further behind as the new information and telecommunications technologies proliferate and become more central to business performance.

Uneven access and capacity underscore the primary challenges rural communities' face in exploiting the new technologies. They must secure cost- and quality-competitive access to advanced telecom services and rapidly build local expertise, training and service capacities to improve local business performance and to attract new firms. The FCC's recent *Report on the Availability of High-Speed and Advanced Telecommunications Services* notes in particular that high-speed telecommunications services are not readily available in rural and low-income areas (FCC, 2000a).[1]

The research strategy used here provides a comprehensive map of current telecom infrastructure and user patterns, information about the effects of access and use barriers on rural businesses, and efforts in Appalachian communities to bridge the digital divide. Specifically, the research plan aimed:

---

[1] It concludes that those outside of population centers are particularly likely to "not be served by market forces" alone.

1. To provide an understanding of the importance of telecommunications producer and user industries for the ARC region as a whole and for urban and rural counties.

2. To provide a profile and analysis of state policies and programs and federally supported investments and incentives to expand telecom access and use in the 13 ARC states.

3. To provide an up-to-date inventory of the telecommunications infrastructure in the ARC region.

4. To provide a detailed examination of access, adoption and implementation barriers in rural communities and highlight successful efforts to overcome these barriers.

For the purposes of this paper, we focus particularly on points 2-3 (the producer and user industries are examined in detail in another paper, Oden and Strover, 2001a). This research presents a snapshot of the fundamental infrastructure issues facing the Appalachian region, and looks at the federal and state universal service programs, as well as state policy and programs addressing infrastructure issues.

II. THE ROLE OF INFORMATION TECHNOLOGIES IN THE ECONOMY

It is widely acknowledged that telecommunications industries that produce information and communications products and services have been a crucial factor in the US economy's sustained and rapid growth during the 1990s. There is a common group of Standard Industrial Classification code jobs that, together, represent aggregate employment in the Information/Communication Technology (ICT) sector. These industries accounted for less then 10 percent of US output during 1995-1999, but close to 30 percent of the country's growth (US Dept. of Commerce, 2000). Employment in these industries grew from 3.9 million in 1992 to 5.2 million in 1998, a 33 percent increase. Similarly, an identifiable group of industries can be coded as telecommunications producing and –using industries. Equipment investments alone nearly doubled, from $243 billion in 1995 to $510 billion in 1999 (US Dept. of Commerce, 2000).

Telecommunications infrastructure is a critical component in these indicators. The presence of and ability to use computers, particularly in a networked environment, and access to appropriate software applications, as well as access to fast communications networks for rapid information flow, are critical to effectively extracting the benefits of information technology. Cronin et al. (1993) found, for example, that telecommunications investment rises with economic growth, while economic growth likewise rises with investment in telecommunications. Parker has reported similar results (1995), as have Dholakia and Harlam (1993). Such data suggest that access to broadband communications networks (200 kbps or faster) at affordable rates will be a significant factor in continued economic growth, as will having the education and training institutions available that can convey to workers the appropriate skills to use network capabilities. This process also underscores that importance of having state and local institutions that can improve access and social capacity to use information and communication technologies in firms, in schools, and in residences.

The federal and numerous state governments have recognized the importance of the communications infrastructural elements that enable economic growth, and they have sought to create an environment that encourages wider deployment of advanced communications capabilities. In the current deregulatory era, this has meant a combination of incentives, government-funded programs, and collaborative ventures with the private sector. For example, the 1996 Telecommunication Act's universal service provisions created an e-rate program to fund Internet connectivity to scholars, libraries and medical facilities. The Rural Utility Service currently grants loans for rural broadband improvements.[2] States, having received much more authority over telecommunications inasmuch as they are the first stop in insuring that the Bell Operating Companies are opening their markets,[3] have sought to create terms and conditions that deliberately encourage statewide network capacity and deployment; several have undertaken assessments of their competitiveness and of their broadband assets. For example, North Carolina completed an exhaustive, exchange-by-exchange study for the entire state inventorying service quality and costs (N. Carolina, 2000). Tennessee's Digital Divide Report includes some data addressing telephone penetration on a county-by-county basis (Tennessee TRA, 2000). Some have sought to use their statewide government communications systems to leverage improved communication services to critical institutions within their boundaries. Several economic development programs at the state level also target telecommunications improvements.

From an industry standpoint, the number of competing local exchange carriers (CLECs) has escalated generally, many of them providing advanced telecommunications services. However, these new competitors face an environment of large and powerful incumbent companies that have lobbied fiercely for fewer restrictions on their abilities to enter new markets, notably long distance voice service and inter-LATA data transport (or long distance backhaul) services.[4] If telecommunications competition in parts of the country seems more intense, in rural regions such as those characteristic of much of Appalachia it seems nonexistent. The limited data that are publicly available demonstrate that broadband deployment is much more widespread and even competitive in populous metropolitan regions, while it is absent in rural America. The National Exchange Carriers Association estimates that it will cost $10.9 billion to make broadband capabilities available throughout rural America (NECA, 2000).

Skirmishes between telecommunications providers, local populations and their officials, and state and federal regulators have broken out over deploying broadband capabilities to rural areas, or even to secondary or tertiary markets.  For example, allowing or encouraging municipally-owned utilities to provide telecommunications services has been the subject of litigation as well as opposing state policies around the country (City of Bristol, Virginia, etc., v. Mark L. Earley, 2001; also Strover and Berquist, 2001).  The process by which Bell Operating Companies open their networks to competitors has proved to be rocky, with several states warning or chastising the BOCs for slow or seemingly deliberate obstructionist behavior (see Pennsylvania's Section 271 deliberations at http://puc.paonline.com/telephone/sec_271.asp).  Portions of the country that want broadband capabilities but cannot obtain them from their local (and usually de facto monopoly) provider have few alternatives.  Satellite broadband systems have been slow to develop; wireless broadband in rural areas have not emerged.

---

[2] The Rural Utilities Service (RUS) announced a new $100 million loan program in December, 2000, that makes funds available to finance the construction and installation of broadband telecommunications services in rural America, targeting through a one-year pilot program broadband service to rural consumers where such service does not currently exist. Communities up to 20,000 inhabitants are eligible.

[3] Under Section 271 of the 1996 Telecommunications Act, state utility commissions must first rules that Bell Operating Companies have opened their market before those companies seek FCC approval to enter long distance markets.

[4] Enabling Bell Operating Companies to provide inter-LATA data transport is one of the key points of the hotly debated Tauzin-Dingell bill in 2001 (H.R. 1542).

Nevertheless, there are some new vendors offering some new services in non-metropolitan markets, and along with them are some new business opportunities.  For example, call centers with their relatively high labor demands have relocated out of metropolitan regions and sometimes find their way to rural areas - as long as the telecommunications infrastructure can support them.  Some communities have made local investment in high-speed networks in order to provide improved services to their businesses.  The "Electronic Villages" of rural Virginia, for example, deliver broadband capabilities to towns with very small populations.

The processes by which high-speed services can be realized in the Appalachian region involve a complex interaction among policymakers, telecommunications companies, local communities, and the local economic environment.  Understanding the volatile climate of lawsuits, evolving policy, and uncertain competitive terrain is a first step to assessing the prospects for broadband capabilities in the region.

III. RESEARCH PLAN

This article reports only a portion of the results of a larger inquiry.  Here, we surveyed secondary data sources and conducted phone interviews with federal and state officials responsible for implementing telecom development programs in the Appalachian Region Commission region to delineate the size and distribution of federal and state telecommunications program in the area.  We also inventoried the regulations and projects pertinent to telecommunications infrastructure developments in the 13 target states, with special attention to: (1) deregulation legislation over the past 5 years; (2) state level competitive assessments or "competition reports" that include data bearing on last mile infrastructure; (3) access (both telephone and Internet) and universal service programs and provisions; (4) agreements to extend service to communities or state and local governments in exchange for state level approval of telecom company mergers; and (5) special state initiatives, including public-private initiatives as well as state networks, that influence the infrastructure (particularly broadband) serving rural areas in particular.

There is no single dataset that compiles a comprehensive and up-to-date listing of state-level telecommunications regulations and related programs.  For the current research, we undertook extensive telephone interviews with key informants (generally agency officials) in each state in order to provide the information on the policies and initiatives noted above as well as web-based and literature searches. Additionally, the FCC along with state regulators established a web-based clearinghouse to serve as a national resource for local communities to share information about their broadband deployment projects, and this site was scrutinized for information (FCC, 2000b).

We gathered information concerning the actual telecommunications infrastructure characteristic of the region. Basic indicators of telecommunications access and use show a wide variation across states in the ARC region. Phone penetration is a common proxy measurement for characterizing telecommunications services and quality in a region, and while the average telephone penetration (phones per capita) in the U.S. is 94%, several of the states in the ARC region fall below that average

The impediments to deploying advanced telecommunications infrastructure – which the FCC has defined as infrastructure delivering at least 200 kbps – include everything from loop lengths, to the nature of a switch at a central office, to perceived demand on the part of vendor companies. The "last mile" infrastructure has been singled out as especially problematic insofar as any individual connection is only as fast as the slowest throughput in its link.[5] The most common technologies to provide advanced telecom links to households and businesses are cable modem and Asynchronous Digital Subscriber Loop or ADSL. However, both have distance constraints.[6]

Assessing last mile infrastructure is a confounding process insofar as the most common geographic unit, the county, has no bearing on the fundamental unit of telephone geography, the exchange (see Nicholas, 2000 and Strover, 1999, for a discussion on this point). Another important geographic unit for telephone systems, the boundaries of the LATA,[7] also can be important in determining the nature of service availability. Basic telecommunications use and access data are available from various sources, although no single source compiles all of the relevant data. We found access to federal level and state data to be problematic in some cases. For example, the FCC's last two datasets on broadband services throughout the country have not been released. Consequently, our analyses using that data are based on reports the FCC received in 1999, sadly out of date for telecommunications infrastructure assessments.

We used two data sources. First, secondary data from various agencies and associations provide a snapshot of relevant capabilities. For example, the FCC maintains a database of central office facilities for the major local exchange companies; the Commission's new Form 477 requires larger providers of local telecommunications and broadband services to report on deployment on a semi-annual basis. The FCC also provides detailed reports on universal service programs. Additionally, states often compile data on providers operating in their region; for example cable television associations operating at the state level generally have some information about their members' facilities so that we can assess cable modem penetration.

---

[5] One can have a fiber-optic based link to a digital switch but if one has a very slow computer, that single element will constrain the speed of the last mile. So too, a poor line connection from a central office to a household or business limits even the fastest computer's ability to enjoy something that looks like an advanced service. The last mile, telephone company vernacular for the connection from customer premises equipment (a home, a business) to a central office, is generally the source of limited bandwidth or speed.
[6] ADSL is generally not feasible beyond 18,000 feet from a central office. Cable television systems serve towns and cities, not truly rural areas.
[7] A LATA is a geographic region denoting a local access and transport area.

## IV. TELECOMMUNICATIONS INFRASTRUCTURE IN THE REGION

Certain sub-areas within the Appalachian states have particularly poor telecommunications infrastructure, a fate of many rural regions around the U.S., while other areas may have excellent capabilities. For example, while North Carolina boasts the Research Triangle with its advanced facilities, it also has five counties that lack any access to high-speed Internet lines (N.C. Dept. of Commerce, 2000), and one quarter of its telephone central offices are in rural counties considered to be economically distressed.[8] Tennessee reports that its own "digital divide" far exceeds the national average (Tennessee Regulatory Authority, 2000). Our research shows that some rural and economically distressed areas do indeed have Internet connectivity and even access to high-speed services. However, the poorest regions of Appalachia seem to lack alternatives, and may pay more for Internet connectivity than their urban counterparts. Table 1 illustrates the state-by-state disparities, and the huge growth rates of the past few years.

**Table 1 Computer, Internet access and telephones**

|  | Percent of Households with Computers | | | Percent of Households with Internet Access | | | Percent of Households with Telephone | | |
|---|---|---|---|---|---|---|---|---|---|
|  | 1998 | 2000 | % Change | 1998 | 2000 | % Change | 1998 | 2000 | % Change |
| AL | 34.3 | 44.2 | 28.9 | 21.6 | 35.5 | 64.4 | 93.3 | 91.9 | -1.5 |
| GA | 35.8 | 47.1 | 31.6 | 23.9 | 38.3 | 60.3 | 91.4 | 91.1 | -0.3 |
| KY | 35.9 | 46.2 | 28.7 | 21.1 | 36.6 | 73.5 | 93.3 | 93.3 | 0.0 |
| MD | 46.3 | 53.7 | 16.0 | 31.0 | 43.8 | 41.3 | 96.5 | 95.0 | -1.6 |
| MS | 25.7 | 37.2 | 44.7 | 13.6 | 26.3 | 93.4 | 89.5 | 89.2 | -0.3 |
| NY | 37.3 | 48.7 | 30.6 | 23.7 | 39.8 | 67.9 | 94.8 | 95.1 | 0.3 |
| NC | 35.0 | 45.3 | 29.4 | 19.9 | 35.3 | 77.4 | 93.1 | 93.9 | 0.9 |
| OH | 40.7 | 49.5 | 21.6 | 24.6 | 40.7 | 65.4 | 95.6 | 94.8 | -0.8 |
| PA | 39.3 | 48.4 | 23.2 | 24.9 | 40.1 | 61.0 | 96.8 | 96.6 | -0.2 |
| SC | 35.7 | 43.3 | 21.3 | 21.4 | 32.0 | 49.5 | 92.9 | 93.2 | 0.3 |
| TN | 37.5 | 45.7 | 21.9 | 21.3 | 36.3 | 70.4 | 94.6 | 95.5 | 1.0 |
| VA | 46.4 | 53.9 | 16.2 | 27.9 | 44.3 | 58.8 | 93.9 | 95.4 | 1.6 |
| WV | 28.3 | 42.8 | 51.2 | 17.6 | 34.3 | 94.9 | 93.8 | 94.0 | 0.2 |
| **U.S.** | **42.1** | **51.0** | **21.1** | **26.2** | **41.5** | **58.4** | **94.1** | **94.4** | **0.3** |

RED: above national average

Sources: NTIA. (July 1999). Falling Through the Net: Defining the Digital Divide; NTIA. (October 2000). Falling Through the Net: Toward Digital Inclusion
FCC. Telephone Subscribership in the US. February 1999 and March 2001

As Table 1 illustrates, all the states joined the national trends toward higher computer penetration and rates of Internet access. Virginia and Maryland stand out with the high penetration rates for both, rates that exceed the national average. We suspect this is due primarily to the intense business development in the Washington D.C. /Maryland area. In 1998, Mississippi and West Virginia were the states with the lowest computer penetration, and both made huge gains from 1990 to 2000 (although they retain the lowest rates among the ARC states even in 2000). Mississippi, West Virginia and North Carolina all had rather low Internet access rates in 1998, but those too increased considerably by 2000, with Mississippi and West Virginia nearly doubling their penetration. Even that growth still left Mississippi with the lowest overall Internet penetration rates, followed closely by South Carolina and West Virginia.

---

[8] High-speed lines were defined very conservatively in this study as 128 kbps for residential service and 256 kbps for business service.

When it comes to the underlying services – the infrastructure – that facilitate access, a picture of very spotty networks and end-user facilities emerges. For example, Figure 1 offers a plot of the locations of fiber backbone points of presence (or PoPs) in the Appalachian region. Traffic in Mississippi and Kentucky faces clear disadvantages since there are few PoPs local to the ARC regions of those states. In Mississippi, for example, data traffic must be hauled either to Tupelo (the location of the marked PoP) or south to Jackson (not in the Appalachian region) or even further north to Tennessee, incurring additional costs. Locations with more PoPs correspond to metropolitan areas as well as to counties along major highways (as in the case of Virginia).

**Figure 1  Broadband PoPs in the ARC Region**

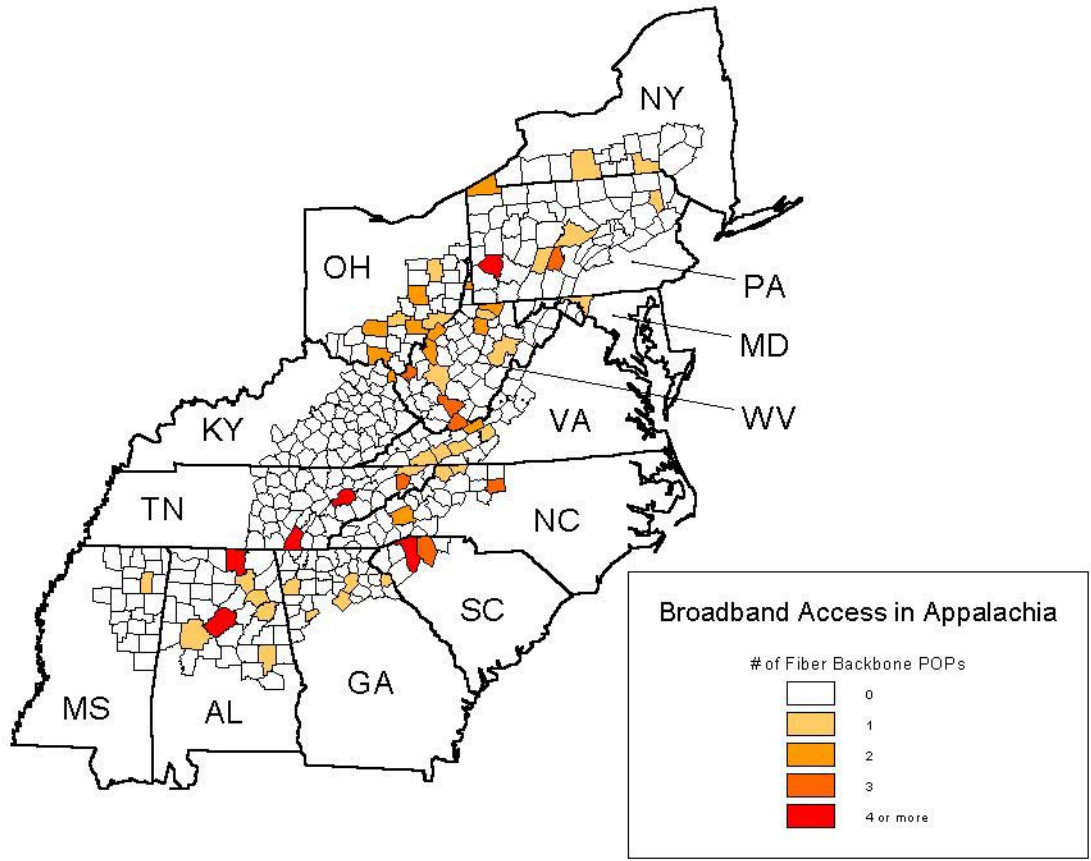

Source:  Authors' telephone conversations with backbone network providers; Boardwatch Magazine's Directory of Internet Service Providers, 13th Edition. (2001).

The PoPs link the "middle mile" and Internet backbone facilities.  The most common high-speed residential and small business end-user technologies are cable modem and DSL services.  When the penetration levels of cable modem and DSL services are examined, we see evidence that these technologies too are underrepresented in the Appalachian Region compared to national averages.  Figure 2 illustrates the locations of cable modem service, although the map is misleading in that it displays the counties where there is cable modem service even though we do not mean to imply that the entire county is actually served.  Cable modem service typically is available only within towns, not in rural areas.  The Appalachian region is sparsely served by this technology, which is confirmed in additional FCC data presented below.

**Figure 2  Cable Modem Service in the ARC Region**

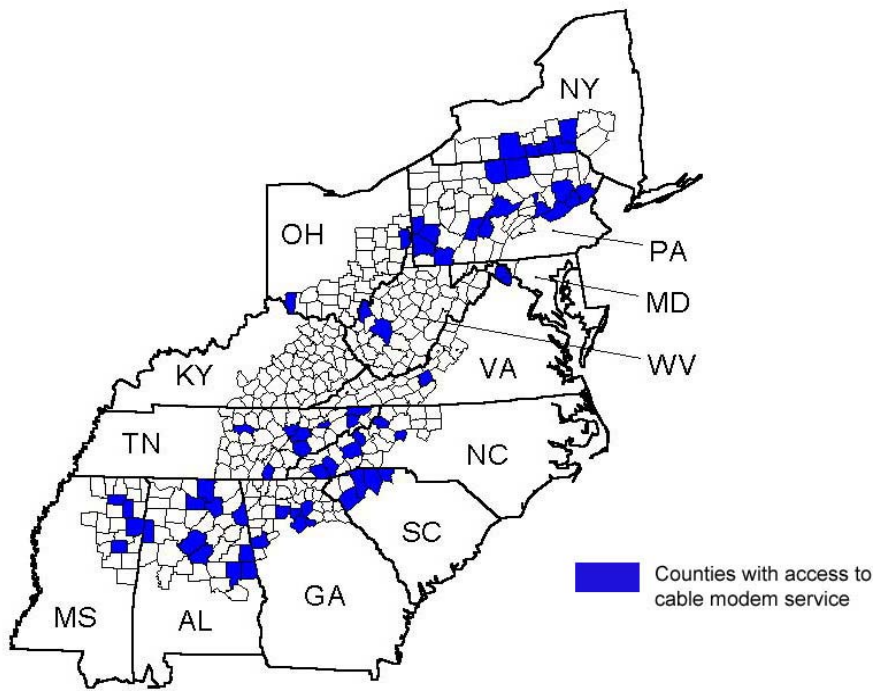

Sources: CableDataCom News. (2001, March 7). Commercial Cable Modem Launches in North America. [Online]. Available: http://www.cabledatacomnews.com/cmic/cmic7.html; Cable Modem Deployment Update. (2000, March). Communications, Engineering and Design (CED) Magazine. M, cited in National Telecommunications and Information Administration & Rural Utilities Service. (2000, April). Advanced Telecommunications in Rural America: The Challenge of Bringing Broadband Service to All Americans. pp. 46-59. [Online]. Available: http://www.ntia.doc.gov/reports/ruralbb42600.pdf

The other major broadband service, DSL, likewise is not broadly available to subscribers in the ARC region.  Kentucky, Ohio, Virginia and West Virginia have a very light presence of  DSL-equipped central offices.  The other ARC states illustrate much broader penetration of DSL-equipped offices.  However, our field visits to Mississippi and Virginia demonstrated that the presence of a DSL-ready office does not necessarily translate into actual DSL service for the region.  For example, the MS counties we visited did not have operational DSL even though Bell South, the dominant local exchange company, said its offices either were or would shortly be equipped for the service and even though those equipped offices appear in public documentation.  We find a statistically significant relationship between the economic vitality of a region (as classified by the ARC as either distressed, transitional, competitive, or in attainment) and numbers of DSL-ready central offices: among the 114 distressed counties, 81% have no

**Figure 3  DSL equipped offices in the ARC Region**

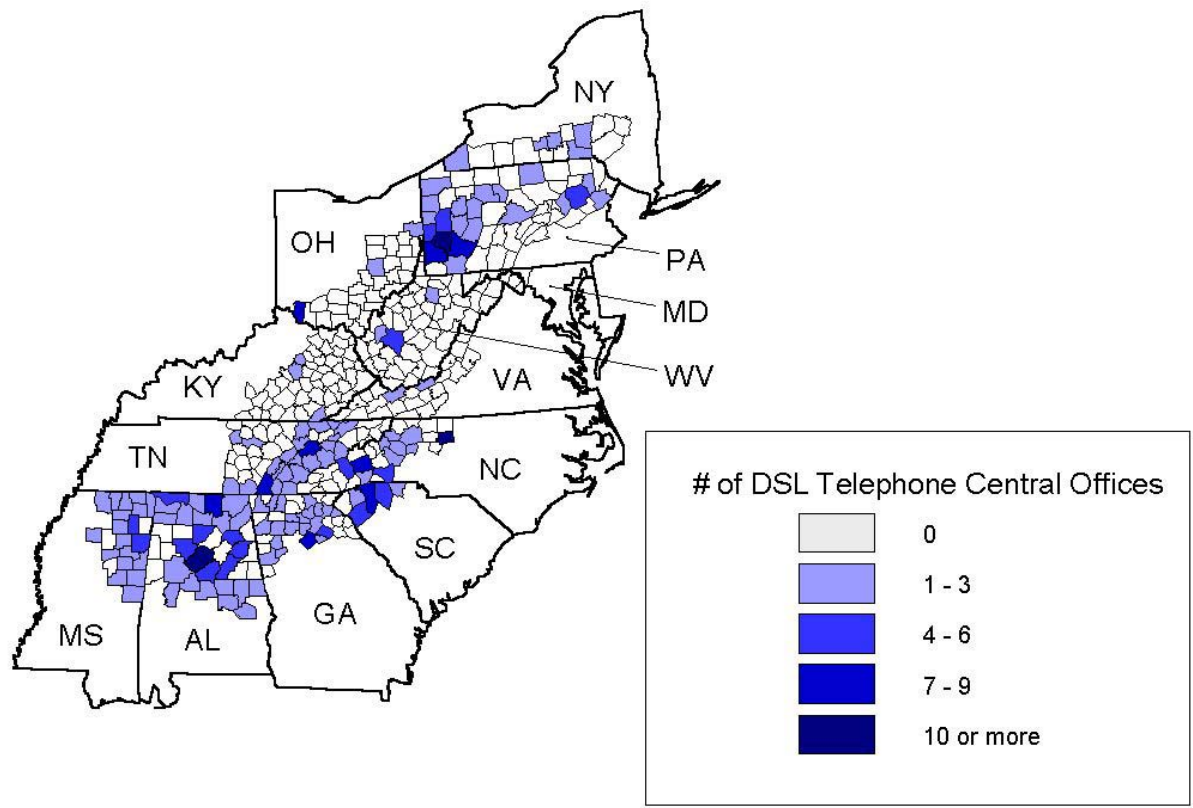

Sources: Authors' search in the Central Office Finder database at DSL
Reports web site. [Online]. Available: http://www.dslreports.com/coinfo;
National Telecommunications and Information Administration & Rural Utilities
Service. (2000, April). Advanced Telecommunications in Rural America: The
Challenge of Bringing BroadbandService to All Americans, pp. 60-72.
[Online]. Available: http://www.ntia.doc.gov/reports/ruralbb42600.pdf

DSL ready central offices, compared to 63% of the transitional counties and 27% of the competitive counties.[9]

The FCC's data from Form 477 categorizes high-speed providers as any service providing at least 200 kbps in at least one direction (user to provider or provider to user).

---

[9] Distressed counties have a 3-year average unemployment rate that is at least 1.5 times the U.S. average of 4.9 percent; have a per capita market income that is less than two-third (67%) of the U.S. average of $21,141 and have a poverty rate that is at least 1.5 times the U.S. average of 13.1 percent OR have two times the poverty rate and qualify on one other indicator.  Appalachian Regional Commission, County Economic Status in the Appalachian Region, FY 2001.

Data they collected illustrate that the more populous regions of Appalachia obtained high-speed services, but many other regions have none. The FCC's use of the high-speed designation is problematic because it does not identify whether the service is broadly available, such as DSL, or a single T-1 line, but in the case of the Appalachian region is it easy to see that high-speed services are not pervasive.

**Figure 4  High-speed Providers**

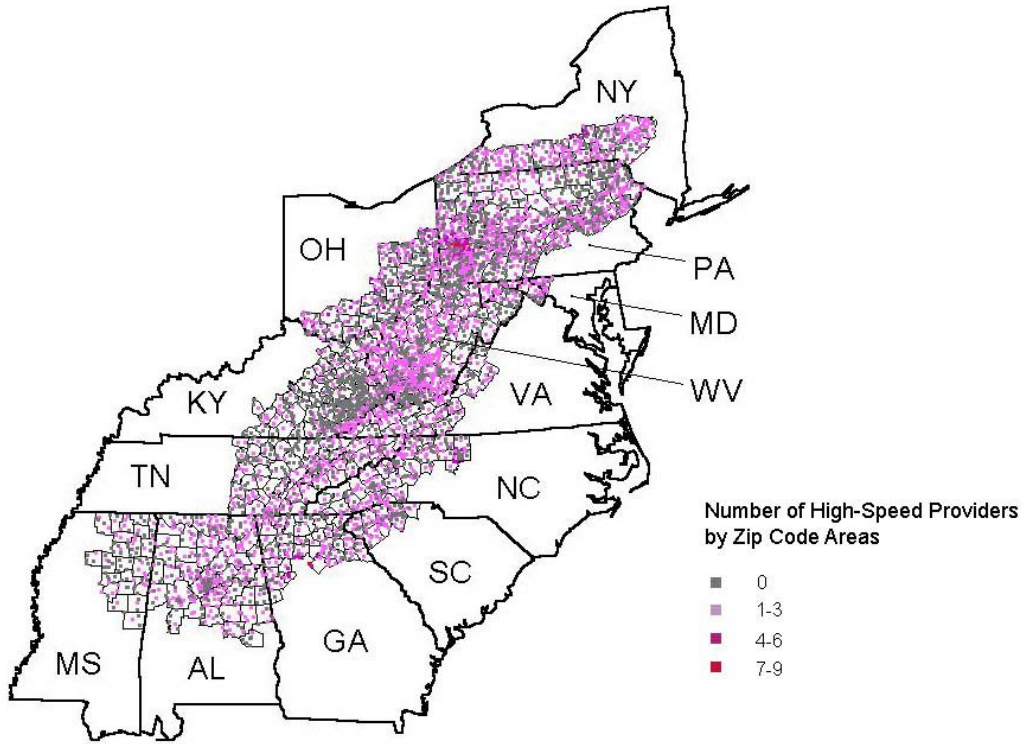

Source: The Federal Communications Commission. (2000, August). Deployment of advanced telecommunications capability: Second report. [Online]. Available: http://www.fcc.gov/broadband/

In fact, we find that 47% of the Appalachian region's zip codes have one or more high-speed service subscribers, compared to the nationwide average of 59% of the country's zip codes, a statistically significant difference. That said, however, the availability of high-speed service can be extremely misleading as an indicator of regional connectivity. In our fieldwork we saw that even in economically distressed counties, the largest businesses had T-1 (or better) connectivity, but that fact said nothing about broader connections and capabilities in the county or zip code. It registers simply as a single line to one business.

Basic line quality and switching features vary tremendously across the Appalachian region, as in other parts of the country. Competitive pressures are relatively low in the Appalachian sub-regions. For example, most of the states with counties in the Appalachian region have fairly low numbers of competing local exchange companies (CLECs), although in two, New York and Pennsylvania, the Bell Operating Companies have been approved to offer long distance services.[10] The December 31, 2000 data as illustrated in Table 2 illustrate that New York has the highest CLEC presence – indeed, it is the highest in the country - followed among Appalachian states by Georgia and Pennsylvania, both with 10% of their end user lines serviced by CLECs.

**Table 2 End user lines (as of Dec., 2000)**

| STATE | ILECs | CLECs | TOTAL LINES | % CLEC SHARE | Bell Operating Companies % of lines, 1999 | Other Price-cap companies % of lines 1999 |
|---|---|---|---|---|---|---|
| AL | 2,351,704 | 191,299 | 2,543,000 | 8 | 79.3 | 12.8 |
| GA | 4,820,788 | 551,316 | 5,372,104 | 10 | 83.3 | 0.6 |
| KY | 2,122,,021 | 56,392 | 2,178,413 | 3 | 56.6 | 34.7 |
| MD | 3,802,622 | 165,502 | 3,968,124 | 4 | 99.8 | 0.0 |
| MS | 1,304,145 | 68,891 | 1,373,036 | 5 | 93.4 | 0.4 |
| NY | 10,962,969 | 2,769,814 | 13,732,783 | 20 | 89.5 | 8.3 |
| NC | 5,071,853 | 286,436 | 5,358,289 | 5 | 50 | 35.9 |
| OH | 6,935,139 | 264,461 | 7,199,600 | 4 | 59 | 33.5 |
| PA | 8,017,391 | 870,618 | 8,888,009 | 10 | 77.1 | 13.1 |
| SC | 2,260,645 | 108,233 | 2,368,878 | 5 | 64.5 | 13.8 |
| TN | 3,291,602 | 296,281 | 3,587,883 | 8 | 79.6 | 10.3 |
| VA | 4,317,626 | 414,432 | 4,732,058 | 9 | 76.2 | 21.3 |
| WV | 927,432 | -- | -- | -- | 83.7 | 14.8 |

Source: FCC, Common Carrier Bureau statistics, 2001.

It is probably safe to predict that the Appalachian sub-regions in these states have lower CLEC activity than that enjoyed by other portions of the state since there are few cities in the those areas. One of the key goals of this research was not only to assess competition but also to assess line quality and upgrade activity in the ARC region. The statistics presented above already point to certain deficiencies in the local and regional networks. The actual cost of providing services in the Appalachian states also is important insofar as longer loop lengths (to serve rural areas, for example) and low population densities mean that those regions should receive more support in order to maintain universal service. Data from the FCC in Table 3 show that five of the 13 states in the Appalachian region have loop costs either at or below the US average of $239 (for 2001).

Alabama, North Carolina, Tennessee, Kentucky, Georgia, South Carolina, West Virginia and Mississippi all have an average loop cost that exceeds the national average cost, and this results in their receiving certain types of universal service support (detailed below).

---

[10] With FCC approval that a state has met its competitive checklist requirements, the BOCs are allowed to enter into lucrative long distance voice and inter-LATA data transport services.

Two of the five states that are below the average, New York and Pennsylvania, have already met Section 271 requirements for competition (the "competitive checklist" constructed by the FCC).

**Table 3 State USF Loops and Loop Costs**

| STATE | USF Loops | USF Cost per Loop | Persons per square mile |
|---|---|---|---|
| Maryland | 3,840,931 | 193.58 | 541.9 |
| Ohio | 7,005,959 | 200.03 | 277.3 |
| Pennsylvania | 8,468,821 | 215.41 | 274.0 |
| New York | 12,818,544 | 220.00 | 401.9 |
| Virginia | 4,762,112 | 239.54 | 178.8 |
| **National Average** | | **239.86** | **80.0** |
| Alabama | 2,521,633 | 272.31 | 87.6 |
| North Carolina | 5,093,322 | 278.71 | 165.2 |
| Tennessee | 3,447,390 | 278.78 | 138.0 |
| Kentucky | 2,191,588 | 298.09 | 101.7 |
| Georgia | 5,208,825 | 304.10 | 141.4 |
| South Carolina | 2,329,487 | 318.00 | 133.2 |
| West Virginia | 1,014,109 | 335.81 | 75.1 |
| Mississippi | 1,420,042 | 352.68 | 60.6 |

Source: NECA's Overview of Universal Service Funds (10/2000)

The FCC recognizes 1301 rural local exchange companies, which serve approximately 6% of US households and cover 35% of the country's landmass, excluding Alaska. These companies typically have longer loops and consequently higher loop costs than companies serving metropolitan regions. However, larger companies including the BOCs, not considered primarily rural telcos, also serve numerous rural households. Bell South, for example, serves most of Mississippi's households. Determining the appropriate amount of support companies serving high cost regions should have in order to maintain the goals of universal service has been a topic of considerable study and lobbying. The FCC adopted a formula for universal service support first for non-rural areas in October 1, 1999[11] and a formula for rural companies in 2001. The impact of that universal service support will be examined below.

## V. UNIVERSAL SERVICE INITIATIVES IN THE APPALACHIAN REGION

A number of federal programs have been initiated to enhance access to basic and advanced telecommunications services, the rationale often (particularly for state funded programs) being to enhance information technology capabilities in rural and low-income urban areas. Here we examine the major federal support program under universal service: the high cost support fund. We also examine several state initiatives as well.

---

[11] High Cost Methodology Order, FCC 99-306.

State programs are a highly heterogeneous collection of endeavors, ranging from leveraging the states' own telecommunications services for broader purposes to operating statewide e-rate-like programs.

A critical question is whether federal funds are distributed evenly among the states, and whether funds are distributed evenly between rural and nonrural areas. In general, the results appear to be mixed, although it is difficult to make the rural/urban assessments simply because certain programs are directed to the entire state rather than specific regions within a state.

Federal Universal Service

While the concept of universal service dates back to the early 1900s, its meaning and mechanisms have undergone several changes.[12] Today, however, federal universal service refers to a series of FCC rules to make various classes of telecommunications services available at just, reasonable, and affordable rates throughout the county, as mandated by Section 254 of the Telecommunications Act of 1996. The current federal universal service policy can be best described as an evolving process, a mixture of formalized regulations, interim regulations, and ongoing debates and proceedings. As such, a thorough description of each component of federal universal service support is beyond the scope of this paper. Instead, we summarize the two key components of federal universal service support—the high-cost program and the E-Rate—and discuss their implications for the particular telecommunications needs of the Appalachian region.[13] Table 4 summarizes the major components of these two programs. Although the 1996 Act does not explicitly state it, the high-cost program goes hand in hand with rate reductions in non-basic services (including long-distance service) so that prices can move toward real costs. Such rate reductions essentially eliminate implicit cross subsidies between non-basic and basic services, one of the goals in the 1996 Act's reformulation of universal service.

---

[12] Mueller, M. (1997). <u>Universal service: Interconnection, competition and monopoly in the making of American telecommunications.</u> Cambridge, MA: MIT Press.

[13] In addition to the high-cost and the E-Rate programs, the federal Universal Service Fund supports low-income programs (i.e., Lifeline and Linkup) and the rural health care program.

**Table 4 Federal Universal Service Components**

| Component | Policy Goals | Mechanism | Implications for Appalachia |
|---|---|---|---|
| High-Cost Program | 1. To prevent the extra cost of providing services in high-cost areas from being reflected on the rates in these areas.<br>2. To create a competitive environment by subsidizing the carriers who serve high-cost areas.<br>3. To transform the historical method of universal service (i.e., implicit cross-subsidies of residential local service by long-distance, business, and non-basic services) to an explicit method in which long-distance, business, and non-basic rates will reflect truer costs. | 1. All telecommunications companies in the country make contributions to the federal Universal Service Fund according to the contribution factor (percentage of end-user revenues), which is decided quarterly by the FCC. Carriers may or may not transfer the burden of contribution to rate payers.<br>2. A portion of the Universal Service Fund is disbursed to eligible *non-rural* carriers that serve high-cost areas. The Fund offsets the cost for the portion that exceeds 135% of national average cost.<br>3. The Fund also supports *rural* providers | 1. Affordable telecommunications services in remote communities that have geographic (e.g., mountains) and economic (e.g., small demands) disadvantages.<br>2. Greater incentives for carriers to enter markets (e.g., rural markets) where service provision is cost prohibitive in the absence of universal service support. |
| E-Rate | 1. To provide access to basic and advanced telecommunications to schools and libraries across the county. | 1. Each school or library that applies to the program receives discounts for connection (e.g., telephone line, T-1, Internet access) and inside wiring.<br>2. The level of discounts ranges from 20% to 90% depending of the economic needs (the number of students eligible for the National Free Lunch Program) and location (i.e., rural or urban). | 1. Affordable access to advanced services (e.g., T-1 connection) in the areas where such services are priced high.<br>2. Make public access terminals with sufficient bandwidth to rural poor residents who lack access at home.<br>3. Greater incentives for carriers to upgrade lines and switches for advanced services (e.g., ATM) because infrastructure upgrades can be partially and indirectly subsidized by the E-Rate when the school or library in one area makes service request for such services. (See the State Network section for detail.) |

In FY2000, the Universal Service Administrative Company distributed $4.4 billion to eligible recipients across the county, and the high-cost program and the E-Rate accounts for the bulk of this support ($4.3 billion) (Universal Service Administrative Company, 2001). The high-cost program (with five separate components) has the largest share in the federal Universal Service, with the amount of disbursement reaching $2.2 billion in FY2000. The E-rate program was the next highest share in the Universal service fund. Of the $2.2 billion high-cost portion of the federal Universal Service Fund (USF), about 30 percent, or $650 million, was distributed to the 13 Appalachian states (Table 5). We can further disaggregate the federal USF by analyzing the relative importance of the federal high-cost program in each of the 13 states.

**Table 5 Distribution of the federal high-cost program in Appalachia in 2000**

| State | Rural$^1$ | Non-Rural$^2$ | Total |
| --- | ---: | ---: | ---: |
| Alabama | $27,833,107 | $60,203,436 | $88,036,543 |
| Georgia | $73,429,979 | $5,919,045 | $79,349,024 |
| Kentucky | $18,839,297 | $10,608,807 | $29,448,104 |
| Maryland | $552,276 | $1,852,272 | $2,404,548 |
| Mississippi | $23,442,921 | $109,658,352 | $133,101,273 |
| New York | $43,566,507 | $9,015,372 | $52,581,879 |
| North Carolina | $24,432,168 | $9,638,988 | $34,071,156 |
| Ohio | $15,579,591 | $3,908,757 | $19,488,348 |
| Pennsylvania | $27,296,823 | $1,459,563 | $28,756,386 |
| South Carolina | $37,895,032 | $11,613,882 | $49,508,914 |
| Tennessee | $29,524,563 | $4,487,319 | $34,011,882 |
| Virginia | $10,656,944 | $26,516,103 | $37,173,047 |
| West Virginia | $25,761,273 | $37,249,836 | $63,011,109 |
| Total | $358,810,481 | $292,131,732 | $650,942,213 |

Source: Universal Service Administrative Company. (2001). 2000 Annual Report: Reaching and connecting Americans, Appendix B. Washington D.C.: Universal Service Administrative Company.
1   "Rural" carriers for the purpose of federal universal service are local exchange carriers that either serve study areas with fewer than 100,000 access lines or have less than 15 percent of their access lines in communities of more than 50,000 in 1996.
2   "Non-rural" carriers are local exchange carriers that do not meet the criteria for "rural" carrier designation.

It must be noted that a precise measurement of the distribution of the federal USF in Appalachia is virtually impossible because of the way the federal USF is disbursed to eligible companies.[14] This is problematic for our purpose because except for West Virginia, all Appalachian states contain areas (i.e., counties) that are not designated as the Appalachia region. For this reason, we will make proxy analyses by focusing on the state-level data.

---

[14] The federal high-cost program disburses the USF to eligible local exchange carriers, but a large number of these eligible carriers have service territories ("study areas") spanning both Appalachian and non-Appalachian counties. Available data from the FCC do not allow us to identify the proportion of universal service support directed to Appalachian and non-Appalachian counties in each state.

**Table 6  Per capita federal high-cost support in Appalachia in 2000**

| State | 2000 Population | Persons per square mile | Total high-cost support | Per capita high-cost support |
|---|---:|---:|---:|---:|
| Mississippi | 2,844,658 | 60.6 | $133,101,273 | $46.79 |
| West Virginia | 1,808,344 | 75.1 | $63,011,109 | $34.84 |
| Alabama | 4,447,100 | 87.6 | $88,036,543 | $19.80 |
| South Carolina | 4,012,012 | 133.2 | $49,508,914 | $12.34 |
| Georgia | 8,186,453 | 141.4 | $79,349,024 | $9.69 |
| Kentucky | 4,041,769 | 101.7 | $29,448,104 | $7.29 |
| Tennessee | 5,689,283 | 138.0 | $34,011,882 | $5.98 |
| Virginia | 7,078,515 | 178.8 | $37,173,047 | $5.25 |
| North Carolina | 8,049,313 | 165.2 | $34,071,156 | $4.23 |
| New York | 18,976,457 | 401.9 | $52,581,879 | $2.77 |
| Pennsylvania | 12,281,054 | 274.0 | $28,756,386 | $2.34 |
| Ohio | 11,353,140 | 277.3 | $19,488,348 | $1.72 |
| Maryland | 5,296,486 | 541.9 | $2,404,548 | $0.45 |
| Nation | 281,421,906 | 80.0 | $2,241,237,733 | $7.96 |

Source: U.S. Census Bureau home page, 2001 Universal Service Administrative Company. (2001).

Table 6 compares the amount of per capita federal high-cost support across the 13 Appalachian states and with the national average.  The amount of per capita high-cost support roughly represents the relative ease of providing basic telecommunications at an affordable and comparable (to urban areas) rate.  There is an inverse relationship between population density (i.e., persons per square mile) and per capita high-cost support.  The amount of per capita high-cost support decreases as the population density increases.

Local service is more costly to provide when there are fewer rate payers and when the rate payers are geographically dispersed.  Indeed, this observation corresponds to the universal service policy goal of the Telecommunications Act of 1996, which underpins federal universal service rules.  The Act attempted to introduce competition to all aspects of telecommunications services, particularly to the local telephone market without sacrificing the affordability of services.  Absent universal service support for carriers that serve high-cost areas, rural telephone markets are not likely to see local telephone competition.

What are the benefits of the federal high-cost program to the 13 Appalachian states? As shown in Table 6, there are six states—Alabama, Georgia, Kentucky, Mississippi, South Carolina, and West Virginia—whose per capita high-cost support either exceeds or is approximately equal to the national average.  These six states are the primary beneficiaries of the federal high-cost program among the 13 Appalachian states.  However, this observation does not reveal whether these six states are taking advantage of the federal universal service in *absolute terms*; a large amount of universal service distribution does not necessarily translate into a positive net inflow-outflow balance between USF distribution and USF contribution.

The current federal USF is designed to create a national pool of funds to which the nation's telecommunications providers make contributions according to the rules set by the FCC.  Once contributions are collected, the USF is distributed to eligible telecommunications carriers (ETCs) to support four universal service programs (i.e., high-cost, E-Rate, low-income, and rural health care programs) without respect to the parity of each ETC's inbound (contribution) and outbound (distribution) flows. In other words, some ETCs contribute more than they receive, while others receive more than they contribute. There are some states that are net contributors and states that are net recipients.

**Table 7  Flow of USF disbursement and contribution in Appalachia in 2000 (in dollars)**

| State | USF Payments to carriers | Contribution to USF | Net flow of funds |
|---|---|---|---|
| Mississippi | 133,052,000 | 18,872,000 | 114,180,000 |
| Alabama | 87,650,000 | 30,116,000 | 57,535,000 |
| West Virginia | 63,061,000 | 12,557,000 | 50,503,000 |
| South Carolina | 50,342,000 | 32,031,000 | 18,312,000 |
| Georgia | 79,527,000 | 72,344,000 | 7,184,000 |
| Kentucky | 29,606,000 | 27,969,000 | 1,637,000 |
| Tennessee | 34,352,000 | 42,882,000 | -8,530,000 |
| Virginia | 37,126,000 | 66,613,000 | -29,487,000 |
| North Carolina | 34,304,000 | 65,174,000 | -30,870,000 |
| Maryland | 2,394,000 | 48,742,000 | -46,348,000 |
| Ohio | 19,587,000 | 76,213,000 | -56,626,000 |
| Pennsylvania | 28,812,000 | 92,096,000 | -63,285,000 |
| New York | 53,021,000 | 159,102,000 | -106,081,000 |

Source: The Federal Communications Commission. (April 2001). State-by-state telephone revenues and universal service data. [Online]. Available: http://www.fcc.gov/Bureaus/Common_Carrier/Reports/FCC-State_Link/lec.html
Note: The figures for the payments from USF to carriers are slightly different from the comparable figures in the total high-cost support column of Table 6 because the figures in Table 7 are rounded up and the two tables are compiled from different source materials. However, the two sets of figures are proximate enough for the purpose of our analysis.

Table 7 shows the balance between USF payments to carriers (inflow) and USF contributions (outflow). A positive number in the net flow of funds for a state means that the state's ETCs receive a greater amount of USF payments than USF contributions made by the telecommunications carriers in the state. That is, those states with positive net flows can be understood as the *true* beneficiaries of the federal high-cost support program. Among the 13 Appalachian states, six states exhibit positive net flows of USF payments and contributions. They are Alabama, Georgia, Kentucky, Mississippi, South Carolina, and West Virginia, and they correspond to the six states that indicate heavy reliance on the high-cost program in Table 6.

Large discrepancies exist among different Appalachian states in terms of both the amount of support flowing into these states and the degree to which they rely on the federal support in maintaining affordable and comparable (to urban areas) rates. Indeed, Mississippi is the country's biggest net recipient of the federal high-cost support while New York is the country's third highest contributor to the federal USF. Strictly speaking, those states that make larger contributions than they receive back are not benefiting from the federal high-cost program. On the other hand, however, the federal high-cost support already has generated positive results among net recipient states.[15]

---

[15] For example, BellSouth in Mississippi received a USF payment in excess of $100 million in 2000, and the company spent the money not only for rate-reduction purposes but also for various infrastructure

State Universal Service

The Telecommunications Act of 1996 allows individual states to implement appropriate support mechanisms for carriers and telephone subscribers to preserve and advance universal service in states.[16] It must be noted that neither the Act nor any other federal laws and regulations require states to create intrastate universal service funds. Therefore, each state must make its own decision as to whether it is appropriate to create an intrastate USF and what is the right size of its USF should be. Such discretion given to individual states has resulted in heterogeneous activities among the 13 Appalachian states (and the rest of the country) in devising and implementing intrastate universal service mechanisms. States opted to established USF programs largely to respond to industry claims for recovering revenue lost due to reduced access rates (and other deregulation initiatives). In this sense, their USF programs have had little to do with responses to citizen needs although in certain states (e.g., North Carolina) some citizens have tried to persuade legislatures to allow community networks to receive universal service funds.

    A. *Commitment by States*

Implementing complicated regulatory mandates demands a tremendous amount of resources, time, and expertise on the part of state regulators. Formulating a universal service policy exemplifies such a case because of the necessity to assemble a complex regulatory mix that calls for complicated cost calculations, associated changes in intrastate and interstate tariffs, consistency with the federal universal service policy, and the requirement to achieve affordable telecommunications rates and competition at the same time. Quite predictably, there is no uniformity in the commitment to the creation of state USFs among the 13 jurisdictions in Appalachia, as shown in Table 8.

**Table 8  Public utilities commission actions for the creation of state USFs in Appalachia  (as of August, 2001)**

| State | USF created or planned[1] | Amount | State | USF created or planned[1] | Amount |
|---|---|---|---|---|---|
| Alabama | No |  | Ohio | No |  |
| Georgia | Yes | $40m | Pennsylvania | Yes | $32m |
| Kentucky | Yes |  | South Carolina | Yes | $41m |
| Mississippi | No |  | Tennessee | Yes |  |
| New York | Yes |  | Virginia | No |  |
| North Carolina | No |  | West Virginia | No |  |

Source: Personal interviews with state public utilities commission staff; the authors' survey of public utilities commissions web sites, the FCC web sites, and general publications.
1 "Planned" means that the state public utilities commission at least has entered an order defining procedural rules toward the creation of a state USF.

---

upgrade projects.  Subsequently, BellSouth's telecommunications infrastructure in Mississippi was enough improved to allow the state government (the Mississippi Department of Information Technology Services) to build a statewide ATM network, which would benefit the state network users (i.e., the state government agencies, local governments, schools, libraries, and universities) by offering greater bandwidth and lower telecommunications costs.  Personal interviews with Gary Rawson (the Mississippi Department of Information Technology Services), Aug. 3, 2001; Randy Tew (the Mississippi Public Utilities Staff), Aug. 3, 2001.

[16] 1996 Telecommunications Act § 254(f).

About half of the 13 states have created or will create in near term state USFs in one form or another.  Recalling the six states that are net beneficiaries of the federal high-cost universal service program (Alabama, Georgia, Kentucky, Mississippi, South Carolina, and West Virginia), there seems to be no relationship between a state's status in federal funding and its commitment to intrastate support.

The universal service policies in four states (Georgia, New York, South Carolina, and Tennessee) were produced through state legislation.  Universal service bills (or telecommunications reform bills that contained universal service requirement) typically define the general policy goals as well as some aspects of implementation procedures for state USFs.  In all four states, state PUCs carried out the actual implementation of USFs.  In contrast to these four states, Kentucky and Pennsylvania's universal service policies were instituted by PUCs and are not codified into their state statutes.

In addition to these six states, the possibility of creating state USFs has been discussed at one point or another by the PUCs in six more states (Maryland, Mississippi, New Carolina, Ohio, Virginia, and West Virginia), but none has initiated specific proceedings for setting procedures and guidelines.

   *B. Types and Nature of Universal Service Support*

The types of universal service support funded by state USFs are varied (Table 9).

**Table 9  USF-supported services in Appalachia**

| State | High-cost | Low-income[1] | Schools/Libraries | Telephone Relay System[2] |
|---|---|---|---|---|
| Georgia | ✓ | | | |
| Kentucky | | ✓ | | |
| New York | | ✓ | | ✓ |
| Pennsylvania | ✓ | ✓ | | |
| South Carolina | ✓ | ✓ | | |
| Tennessee | ✓ | ✓ | ✓ | ✓ |

Source: Personal interviews with state public utilities commissions staff; the authors' survey of public utilities commissions web sites, the FCC web sites, and general publications.
Note: All 13 Appalachian states but Ohio provide low-income support at a state level, but those states that are not listed on the table have not created explicit USFs.
1  Low-income support includes Lifeline and/or Linkup, and the state low-income support supplements the federal low-income universal service support.
2  Telephone relay system is a service for people with hearing disability.

There is no discernable pattern among the universal service policies among the six states.  Each has a unique combination of USF-supported services, but high-cost support and low-income support are the most popular types of intrastate USF support.  The high-cost component is the most prominent aspect of a state USF from a regulatory perspective because low-income support by states is a relatively *passive* policy measure, while the high-cost component of state USFs is an *active* policy measure.  We can understand this difference by considering the relationship between the federal and state USFs.

The Lifeline portion of the low-income customer USF support has a three-part design. The federal USF provides a baseline assistance of $6.1/monhth/line to all 51 states. States then decide whether to provide additional support (up to $3.5/month/line). For those states that provide additional Lifeline assistance, the federal USF provides a matching support (1/2 of the amount of state-level support). In contrast, there is no regulatory mechanism or requirement to coordinate state and federal high-cost support. The creation of a state USF with high-cost component is completely a discretionary activity of each state. In this respect, the four states with their own high-cost programs (Georgia, Pennsylvania, South Carolina, and Tennessee) are arguably the most active states in Appalachia in terms of universal service policy.

A state's decision to create a USF may not be directly contingent on the size of its federal USF distribution. This point may be supported by the fact that Pennsylvania, whose federal high-cost support for non-rural carriers is considerably smaller than the support the state's rural carriers receive (see Table 5), excludes Verizon (a non-rural carrier) from the group of USF eligible carriers. That is, Pennsylvania's state USF is not designed to compensate for the shortage of federal support to non-rural carriers, but rather its goal is to further increase the support for rural carriers, which are already receiving a larger federal USF disbursement than non-rural carriers.

## VI. STATE AND LOCAL INITIATIVES

Ever since AT&T's divestiture became effective in 1984, state legislatures and their utilities commissions have had much more responsibility for monitoring and regulating telecommunications activities in their boundaries. Each of the Appalachian states has chosen distinctive paths to handle its regulatory responsibilities. Some appear to have much closer relationships to large, incumbent companies than others; some have considerable staff resources and expertise to help establish policy, while others, such as Mississippi with its two telecommunications staff people, have very limited resources. In this section we investigate the range and depth of various programs and policies states have adopted to enhance the delivery of telecommunications services. Some mechanisms include using state networks to enhance non-state communications opportunities, using utility commission approval over mergers or 271 proceedings to leverage concessions from carriers, establishing special programs targeting rural digital inequities, and establishing unique joint ventures with carriers in order to achieve improved statewide infrastructure. Certain cities and towns also have initiated telecommunications projects to enhance local connectivity and opportunities for economic development.

### State Networks

So far we have discussed the status of telecommunications infrastructures in the Appalachian region as developed primarily by private telecommunications companies for private profit. However, the state governments of the 13 Appalachian states have developed numerous infrastructure and connectivity projects over the years. These projects are pegged on wide-ranging goals ranging from simply making telecommunications bills to state agencies cheaper to boosting public telecommunications infrastructure upgrading throughout the state to the benefit of state government, business users, and the general public. Accordingly, the technological underpinnings and mechanisms of state telecommunications networks tremendously vary from one state to another. Our observation of different state networks in Appalachian states illuminates some characteristic features that are common to several state networks. In this section, we offer a typology of state telecommunications networks in the 13 Appalachian states, and make an assessment of the impacts of each network type on the overall connectivity and access to advanced telecommunications technologies in these states.

**Table 10  Typology of State Telecommunications Networks**

|  | Goals | Mechanism | Adopted in |
|---|---|---|---|
| **a. Demand Aggregation** | 3. To lower telecommunications costs for the state and other government users. | The state government receives volume discounts from telecos by consolidating telecommunications service demands of various state government agencies and offices into a single large purchasing unit . | 4. Virginia |
| **b. Resource-Sharing** | 4. To lower telecommunications costs for the sate and other government users. <br> 3. To maximize the efficiency of existing and new telecommunications infrastructures in key routes. | The state government and a telco barter free access to the state's highway rights of way and free telecommunications services to the state government and/or telecommunications infrastructure ownership. The state government and the vendor usually make a commitment to a long-term partnerships that may last for several decades. | 2. Maryland <br> 3. New York <br> 4. South Carolina <br> 4. West Virginia |
| **c. Anchor Tenancy** | 1. To lower telecommunications costs for the sate and other government users <br> 1. To upgrade public telecommunications infrastructure in all parts of the state. | The state government and a telco or telcos enter a contract to make advanced telecommunications available to the state government. Telecommunications service to the state government is provided through public telecommunications networks, which would receive switching and transport capability upgrading as specified in the contract. Such an infrastructure improvement benefits all telecommunications users in the state (i.e., businesses and residents) because public telecommunications networks are used by all types of users. | 1. Alabama <br> 2. Georgia <br> 3. Kentucky <br> 4. Mississippi <br> 5. New York <br> 6. North Carolina <br> 7. Ohio <br> 8. Pennsylvania <br> 9. Tennessee |

Table 10summarizes our observations of state telecommunications networks in the 13 states. We have identified three major types of networks.

   *Demand Aggregation*

A state telecommunication network under the demand aggregation model creates a single large telecommunications customer by consolidating telecommunications service purchases of state agencies and other eligible users (e.g., public schools and libraries, local governments, universities and colleges, etc). The advantage of the demand aggregation strategy is the cheaper telecommunications services for participating users. The model leverages the state government to negotiate services prices that dispersed, individual users could not receive.

The Commonwealth of Virginia Network (COVANET) exemplifies the demand aggregation strategy. COVANET, like many other state telecommunications networks, is built upon the public telecommunications infrastructure, owned and operated by a private telecommunications company (MCI Worldcom). Prior to the COVANET contract in 2000, at least five separate state-funded telecommunications systems had existed in Virginia. The principal goal of COVANET was to consolidate the voice, data, and video transmission requirements interspersed among existing, separate networks into a unified system. As a result of the network and demand consolidation, COVANET succeeded in substantially lowering various telecommunications rates for the public sector users, including schools and local government.[17]

   *Resource Sharing*

State governments maintain relatively few assets at their disposal that they can turn into economic gains to the state governments. However, states can potentially gain desired outcomes for economic development strategy and meet their own telecommunications needs by taking advantage of their ownership and control of highway rights-of-way (ROW).

Different states have different policies in authorizing the use of highway ROW; some states grant private telecommunications companies the right to lay telephone and fiber lines on highway ROW free of charge, while others demand monetary compensation from telecommunications companies. Under a resource-sharing arrangement, the state government and a telecommunications company "barter" the company's free access to highway ROW and the state government's access to and/or ownership of a portion of telecommunications facilities developed under the resource-sharing arrangement. The primary benefit of the resource-sharing model to states is that the state government gains access to new (typically fiber optic) infrastructure without using any public money. The model also allows the state government to create incentives for infrastructure expansion to its partner (i.e., the telecommunications company under the contract). A resource-sharing agreement is typically a long-term contract that extends over a few decades. Thus, the telecommunications company under the contract is assured of the predictability and stability. The downside of the resource-sharing model is the problem of technological obsolescence. Because the contract locks the state into the types of technologies and services specified in the contract, the state may not be able to adopt future new technologies without bearing additional costs. In addition, the model is discriminatory by its nature, limiting the contract's economic opportunities to one or a small number of select companies.

---

[17] COVANET reduces voice long-distance service by 32-52 percent, T-1 Frame Relay by 20-28 percent in comparison to pre-2000. COVANET also reduces ATM rates by 15 percent from the rates the state government received under the Net.Work.Virginia. deal. (Net.Work.Virgini. is a consortium lead by Verizon and Sprint and provides advanced telecommunications services at discount prices to Virginia's public and private entities. Since the creation of COVANET, the state government encouraged state agencies and schools to switch from Net.Work.Virginia. to COVANET.)  See, Carter, L. & Davidson, B. (2000, May 10). Covanet. Presentation given at the Customer Summit, Virginia Department of Information Technology. [Online]. Available: http://www.dit.state.va.us/telco/covanet/

Several states in Appalachia have adopted the resource-sharing strategy to develop their state telecommunications networks.[18] Maryland is one example, with its innovative approach in linking the resource-sharing model to the development of a statewide telecommunications network. Unlike many other states, Maryland has designed its state network, Net.Work.Maryland, on newly constructed infrastructures. In order to alleviate the enormous costs associated with such a ground-up project, the state government entered into three resource-sharing contacts with private telecommunications companies. Net.Work.Maryland—currently being built in several phases with a schedule of first service delivery in October, 2001 in limited areas—is envisioned as a
network to provide state of the art telecommunications service to Maryland's state agencies, local governments, educational institutions, health care facilities, and, quite notably, private businesses.[19]

Maryland has entered three separate resource-sharing contracts with four telecommunications companies.
1. A 40-year contract with MCI Worldcom and Teleport Communication Group. The companies provide 75 miles of fiber optics and the services required to activate the fiber along the Baltimore/Washington corridor. The state receives free bandwidth service (OC-48). The estimated value of the contract is $32.8 million.
2. A 40-year contract with Level 3 Communications. Level 3 provides 330 miles of fiber optics from the southern portion of the state to the central, east, and west portions. The estimated value of the contract to the state is $222.8 million.
3. A 10-year contract with Willimas Communication. The company provides fiber lines and related equipment in the Baltimore area. The estimated value of the contract to the state is $9.4 million (Association of Telecommunications Professionals in State Government, 2000).

Together, these resource-sharing contracts will form a 13-node high-capacity

---

[18] Maryland, New York, South Carolina, and West Virgini.
[19] A legal ambiguity remains with regard to the access to Net.Work.Maryland by private businesses. The Task Force on High Speed Networks strongly recommends such access. See, Task Force on High Speed Networks. (1999, December 31). Report to the Maryland General Assembly. [Online]. Available: http://www.techmd.state.md.us/Technology/TFHSN/leg-report.pdf

(10 Gbps) fiber optic backbone connecting all four LATAs. Each county will have at least one POP (at least 45 Mbps), which will be supplemented by 133 fiber "drops" at various highway intersections for future connections.

*Anchor Tenancy*

The third model of state telecommunications network strategy is the anchor tenancy model, and this model has been most widely adopted by the states in the Appalachian region.[20] Like the demand aggregation model, the anchor tenancy model is characterized by the reliance on public telecommunications infrastructure owned by private telecommunications companies. Indeed, the difference between the two models is small, yet explicit. In the anchor tenancy model, the state government becomes the principal (anchor) tenant of a private telecommunications company's public network, guaranteeing a certain level of service purchase (i.e., demand aggregation). In turn, the contract requires the telecommunications company to make a commitment to infrastructure upgrading and service deployment as requested by the state government. The key to understand the benefit of state network is the use of public telecommunications infrastructure. The state makes the request to the telecommunications company under the contract to make infrastructure upgrades to meet the telecommunications needs of the state government (and other eligible users). However, since the state network is built upon leased public telecommunications infrastructure, the improvement in the technological capabilities in the public system benefits all telecommunications users who share the same system.

The experience of Mississippi illustrates the anchor tenancy model mechanisms. Started as a Frame Relay network in 1995, the State of Mississippi revamped the network in 2000 and converted it into the Statewide ATM Backbone Network. The Statewide network consists of seven ATM nodes or "clouds" located in Jackson, Greenwood, Tupelo, Meridian, and Hattiesburg within the state's primary LATA, and in Memphis and Gulfport to serve the Northwest Mississippi LATA and the South Mississippi LATA. Although the majority of these nodes are located outside the Appalachian portion of Mississippi, the contract explicitly requires the contractor (BellSouth) to make necessary upgrades in all the state's counties including the 22 Appalachian counties in order to bring ATM access to all counties. Each user of the Statewide Network is responsible for furnishing the "last-mile" (typically a T-1 connection) connection between the user site and the nearest BellSouth telephone central office, but the contract requires the connection to be a non-mileage sensitive rate that is uniform across the state. The Statewide Network is open to state agencies, universities, community colleges, K-12 schools, public libraries, and local governments.

Mississippi's telecommunications infrastructure has lagged behind the nation for years because of the rural nature of the state.[21] The anchor tenancy model adopted by Mississippi envisions radically changing the infrastructure capabilities of Mississippi's public telecommunications infrastructure since the Statewide Network project requires the contracting telecommunications company to convert most of wire centers into digital systems to transport data to ATM nodes. Such improvement at the wire center level benefits businesses and other telecommunications users because wire center facilities contain a lot of shared equipment that equally benefit the Statewide Network users and other users. Thus, the anchor tenancy model in Mississippi is designed to bring technologies and services to areas that would otherwise be considered "uneconomical" markets that do not justify upgrading.

---

[20] These three models should not be considered mutually exclusive. The State of New York, for example, employs both resource sharing and anchor tenancy models for the creation of its state telecommunications network.

[21] In 2000, the Mississippi's federal high-cost support was the highest.

Utility Commission Authority

Of all the Appalachian states, New York has sought an orderly and monitored deregulation program most aggressively. It began deregulating its local exchange companies in 1985, well before the 1996 Telecommunications Act was passed. In 1995 it opened local exchange markets to competition, and undertook a variety of price controls, gradually lifted, in order to grow competition in the state. Its Public Service Commission required Bell-Atlantic to commit to a one billion dollar infrastructure upgrade program in 1997 as part of its approval of that company's merger with Nynex, and the commission was one of the first to initiate a rigorous review of Verizon as it sought approval under the Section 271 requirements.

Ohio and Pennsylvania also have taken advantage of occasions requiring merger approval to stipulate new or improved services from telephone companies. In approving the Bell-Atlantic-GTE (Verizon) merger in 1999, the Pennsylvania PUC required that the new company provide broadband capability to 50% of the state by 2004 and to the rest of the state by 2015, with the proviso that deployment to balanced across urban, suburban and rural areas. In addition to its active role taken in the Bell-Atlantic-GTE merger proceeding, the Pennsylvania PUC stands out in the crowd by having attempted to restructure its dominant Bell provider. In March 2001, the Pennsylvania PUC entered an order demanding the functional structural separation of Verizon into retail and wholesale units.[22] The goal of the structural separation of Bell companies is to remove barriers for local telephone competition by "structurally" preventing Bell companies from favoring their own local telephone services over those of competing local exchange carriers who lease Bell local facilities. Although the functional separation order was rescinded, this effort indicates Pennsylvania's active commitment to the creation of competitive telecommunications markets. Ohio also required the newly merged Ameritech-SBC to deploy DSL in both rural and urban areas in its 1999 merger approval. It also required a $2.25 million fund to assist rural and low-income areas in accessing advanced telecommunications technology. Virginia also established several conditions for approval the Bell-Atlantic-GTE merger, including infrastructure and service upgrade requirements (1999).

 Special Programs

---

[22] The PUC originally sought a full structural separation of Verizon into two independent companies. See, Global Telephone Order (1999, September 30).

Georgia stands out as a state that enabled municipal governments to be eligible for local exchange carrier licenses as early as 1995. The Governor of Georgia announced a rural broadband initiative in May 2000, which promises to bring broadband infrastructure to rural regions. The network will support download speeds of 1.5Mbps. Another more modest project in collaboration with BellSouth will connect all K-12 school districts to the Internet with a T-1 connection; all Georgia's Appalachian counties are scheduled to receive these connections. As recently as summer, 2001, the state approved a novel non-profit consortium of 31 towns and cities and one county to offer broadband telecommunications services in a wide variety of locations (GeorgiaPublicWeb, 2001).

Maryland, dominated by Verizon, has several programs to encourage e-commerce and an overall statewide information technology program. In 1998 it instituted a property tax credit (HB 477) that awards commercial and residential tax credits for renovations to accommodate advanced computer and telecommunications systems. Additionally, there are two investment funds to support innovative technology efforts in the state. Its key architecture, however, is its statewide network plan to have a point of presence in all three Appalachian counties and to link communities via high-speed fiber.

Virginia has a unique resource in the form of Virginia Tech University, which has purchased four wireless spectrum licenses in the rural western portion of the state in order to experiment with alternative broadband services. This University also has spearheaded several "electronic village" initiatives. North Carolina and Tennessee are notable for having studied their infrastructure characteristics; in the case of North Carolina, a detailed exchange-by-exchange investigation was undertaken. Ohio also undertook an infrastructure assessment under the auspices of the Ecom-Ohio effort centered at Ohio State University.

Tennessee is one of the seven Appalachian states that approved municipally owned utilities providing telecommunications services (in 1997). The others include Kentucky, North Carolina, Alabama, South Carolina, West Virginia and Georgia. The Electric Power board of Chattanooga was the first municipal utility to be certificated for telecommunications services under that law, and it serves five counties in the Appalachian region.

In addition to sponsoring a statewide network that is available to non-government users, North Carolina also passed a bill in 2000 to create a new state agency charged with overseeing rural economic development and information technology infrastructure in the state. This agency is to serve as a rural Internet access planning body, and has as its goal ensuring that dial-up access is available in every exchange by the end of 2001, and that high-speed Internet access is available by 2003 to all citizens of the state.

West Virginia, Mississippi, South Carolina and Alabama stand out as a handful of states that have sponsored or pursued few initiatives to aggressively enhance their telecommunications infrastructure.

Local Initiatives

Several towns and cities in the Appalachian region have initiated efforts to develop local advanced telecommunications services. They include Calhoun, GA, Abington, VA, Blacksburg, VA, other "electronic villages" in rural western Virginia, as well as the notable challenge by the City of Bristol to Virginia's prohibition on municipally owned telecommunications operations (Neidigh, 2001). Such innovations are notable in that in most cases (Bristol being the exception), local exchange carriers either aided the towns' efforts or at minimum did not challenge them. They also are notable in that local leaders believed that telecommunications capabilities would substantially enhance their economic base, either by servicing existing businesses or by attracting new businesses.

Bristol's fiber optic network, begin in 1999, allows the utility to manage load requirements and also deliver services for Internet, LAN extensions, telephone and video conferencing, and virtual private networking to schools and government offices. The utility is moving toward an open access network that would allow non-facilities-based providers access to the network in order to broaden the service base to residential and commercial customers at a competitive price. Its backbone now consists of 125 miles of 144 and 288 count cables, supporting an ATM network operating at 622 Mbps. It is expanding to accommodate gigabit Ethernet, and supports nine points of presence providing collocation facilities. The effort ran afoul of a state law (HB 335) passed in 1998 that prohibited any locality in the state from establishing a "governmental entity" having the authority to provide telecommunications services. Notably, the town of Abingdon was explicitly exempted from this law. Bristol challenged the law and won its case in federal court. However, the case is on appeal at this writing.

In North Carolina, all 29 Appalachian counties had participated in Connect NC project between 1996 and 1999. Connect NC was an educational campaign targeting the public and private sector leaders in the rural western parts of North Carolina with information regarding the importance of telecommunications connectivity to the economic competitiveness of rural areas. Western North Carolina was divided into six regions and each region developed and pursued various pilot projects with a goal to enhance telecommunications connectivity. A notable effort came from a group composed of Alexander, Burke, and Caldwell counties, which created regional WANs to bring high-speed Internet connection at T-1 speed (1.5 Mbps) to municipal and county facilities including public terminals at libraries.

Another innovative wireless Internet project was implemented in Ohio. Sequelle is a non-profit corporation created by Washington county Community Improvement Corporation to provide broadband wireless Internet access and support services to Southeastern Ohio and Northwestern West Virginia at an affordable price. The project is the first in its kind in the nation and aims at promoting economic and community development in rural areas where advanced telecommunication technology is lacking. Sequelle uses the two-way digital FCC-licensed radio frequencies. It is estimated that Sequelle will have 300 customers by the end of the third year (2002). The projected service cost is about 40% of the cost of comparable commercial offerings. The service was initially rolled out in Washington County (OH) and Wood County (WV). The project is estimated to cost $3 million, and is funded by a combination of state and federal funds. The project is designed to become self-sustaining within a few years.

Table 11 lists the local initiatives captured by two federal programs from NTIA and the Department of Education as well as those entered into a community-based database, CTCNet.

**Table 11  Number of community technology centers in Appalachia**

| State | Number of CTC sites in Appalachia |
| --- | --- |
| Pennsylvania | 65 |
| Alabama | 25 |
| West Virginia | 14 |
| Tennessee | 13 |
| North Carolina | 12 |
| Kentucky | 11 |
| New York | 11 |
| Ohio | 7 |
| Mississippi | 3 |
| Virginia | 3 |
| Georgia | 2 |
| Maryland | 2 |
| South Carolina | n/a |

Sources: The Neighborhood Network page in the U.S. Department of Housing and Urban Development web site: http://www.hud.gov/nnw/nnwindex.html; the Technology Opportunities Program database in the National National Telecommunications and Information Administration web site: http://www.ntia.doc.gov/otiahome/top/grants/search.htm; the community technology center database in the Community Technology Center's Network web site: http://www2.ctcnet.org/ctc.asp

## VIII.  CONCLUSIONS

This research has sought to document the status of telecommunications in the Appalachian region with a view to assessing its potential relationship to economic growth and the range of federal and state policies that influence its development.  We have found that telecommunications infrastructure in the Appalachian regions is less developed than that in other parts of the country and that it compares negatively to national averages. Broadband technologies such as cable modems, DSL, and even the presence of high-speed services are not as widely distributed in our target region as national statistics would suggest.  Statistical analyses show that these distribution patterns are in each case associated with economic activity:  more distressed counties have less developed broadband telecommunications infrastructure.

We also find that federal universal service benefits are limited to the most rural of the Appalachian states: only Mississippi, Alabama, West Virginia, South Carolina, Georgia and Kentucky have a net positive inflow of funds through the program, although the internal adjustments (from larger, urban-serving companies to smaller, rural companies) among the other states are not to be discounted.  These six states are among the most rural of all the Appalachian states, having the lowest population densities among the group we are examining (Tennessee being a close exception).

While state universal service programs have cropped up in part to ameliorate the revenues losses local exchange companies face attribute to deregulation (especially reduced access rates), those programs are not uniform. Most offer some low-income support as well as support to telecommunications companies serving high cost territories. Some states are not allowing that support to flow to the largest, wealthiest companies (e.g., the BOCs or other price-cap companies) and instead favor companies serving exclusively rural regions. In such approaches they hint at the sorts of concerns for balancing costs and supports that will probably become more pervasive in the future.

Several states have proactively initiated programs to enhance telecommunications infrastructure. By using state telecommunications networks through resource sharing, demand aggregation or anchor tenancy programs, states are able to leverage their considerable investment and offer benefits to other public sector users – and in some cases, even private sector users. Seven states also allow municipally owned utilities to offer telecommunications services, expanding the range of choices and the potential for competition in the process. Nearly every state had some special program, or many programs, for enhancing Internet connectivity or broadband access. North Carolina and Tennessee for example undertook studies of the situation in their regions. Georgia initiated an aggressive broadband initiative that is supposed to expand access to the entire state quickly. Several states prioritized inexpensive and fast networks serving educational institutions and libraries so that these critical mediating institutions are well served. The least active states appear to be West Virginia, Mississippi, South Carolina, Kentucky and Alabama, although these too have some state programs to enhance telecommunications access or use.

One factor that appears to enhance state potentials for improved telecommunications is coordination among state agencies within the state. By coordinating network design and use, state-funded infrastructure can be used optimally. When it is absent, programs may be duplicative, underutilized, and more costly.

Most state and federal programs have focused on market-related initiatives to solve their telecommunications problems. We observe, however, that attempting to work with (or against) the market yields only limited returns in the absence of leadership. With more creative collaboration and attention to some of the nonmarket solutions to obtaining and using telecommunications - solutions such as training, education, organizational resource sharing – the larger harnessing of telecommunications capabilities to economic growth can be enhanced.